\documentclass[preprint,prc,showpacs,preprintnumbers,amsmath,amssymb,showkeys,floatfix,superscriptaddress]{revtex4}

\usepackage{graphicx}
\usepackage{dcolumn}
\usepackage{bm}

\begin{document}


\title{Investigation of photoneutron reactions close to and above the
neutron emission threshold in the rare earth region}

\author{J.\ Hasper}
\affiliation{ Institut f\"{u}r Kernphysik, Technische
Universit\"{a}t Darmstadt,
 Schlossgartenstra\ss e 9, D-64289 Darmstadt}
\author{S.\ M\"{u}ller}
\affiliation{ Institut f\"{u}r Kernphysik, Technische
Universit\"{a}t Darmstadt,
 Schlossgartenstra\ss e 9, D-64289 Darmstadt}
\author{D.\ Savran}
\affiliation{ Institut f\"{u}r Kernphysik, Technische
Universit\"{a}t Darmstadt,
 Schlossgartenstra\ss e 9, D-64289 Darmstadt}
\author{L.\ Schnorrenberger}
\affiliation{ Institut f\"{u}r Kernphysik, Technische
Universit\"{a}t Darmstadt,
 Schlossgartenstra\ss e 9, D-64289 Darmstadt}
\author{K.\ Sonnabend}
\affiliation{ Institut f\"{u}r Kernphysik, Technische
Universit\"{a}t Darmstadt,
 Schlossgartenstra\ss e 9, D-64289 Darmstadt}
\author{A.\ Zilges}
\affiliation{ Institut f\"{u}r Kernphysik, Technische
Universit\"{a}t Darmstadt,
 Schlossgartenstra\ss e 9, D-64289 Darmstadt}
\affiliation{ Institut f\"{u}r Kernphysik, Universit\"{a}t zu
K\"{o}ln, Z\"{u}lpicher Stra\ss e 77, D-50937 K\"{o}ln}

\date{\today}

\begin{abstract}
We have investigated the photoneutron cross section of the
isotopes $^{148,150}$Nd, $^{154}$Sm, and $^{154,160}$Gd close to
the neutron emission threshold in photoactivation experiments at
the Darmstadt superconducting electron linear accelerator
S-DALINAC. Naturally composed targets were activated with a
high-intensity bremsstrahlung beam at various energies and the
reaction yields have been determined by measuring the activity of
the produced radioactive isotopes with HPGe detectors. The results
are compared to two different statistical model calculations.
\end{abstract}

\maketitle

\section{Introduction}
\begin{figure}[htbp]
  \includegraphics[width=1.\columnwidth]{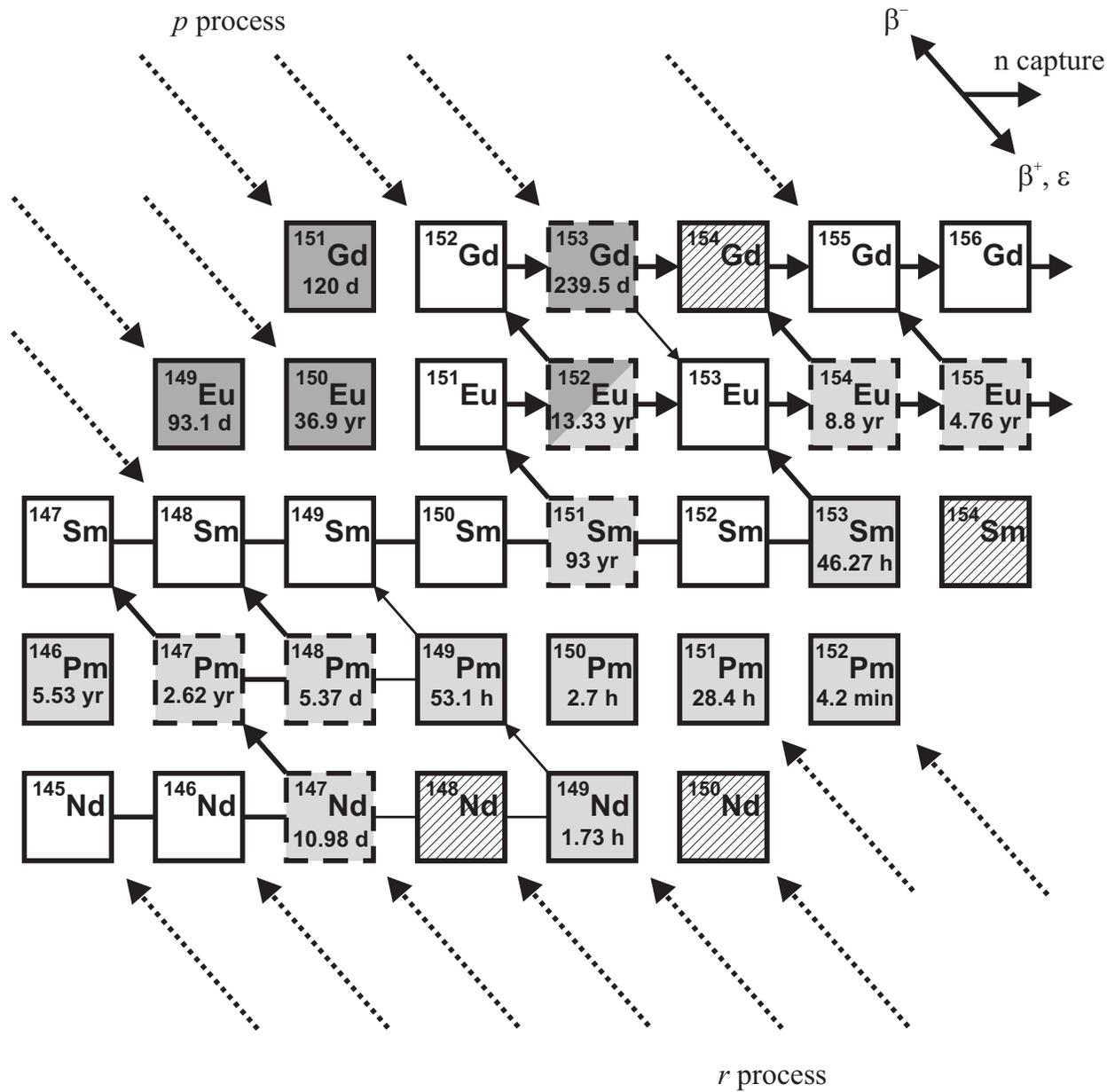}\\
  \caption{The \emph{s}-process flow (solid lines) in the rare earth
  region. The thickness of the solid lines indicates the strength
  of the reaction flow. The reaction path is influenced by several branching points (dashed boxes). Unstable isotopes are shaded light
  grey ($\beta^-$) and dark grey ($\beta^+,\epsilon$). Most isotopes also have nucleosynthesis contributions from the \emph{r}
  and \emph{p} process (dotted lines), respectively. The photoneutron reactions of the hatched isotopes have been studied in this work.}\label{fig:process_flow}
\end{figure}

The nucleosynthesis of the elements heavier than iron is dominated
by the slow neutron capture process (\emph{s}~process) and the
rapid neutron capture process (\emph{r}~process). However, 35
proton-rich stable isotopes cannot be produced in either of these
two processes and are believed to be mainly synthesized by a
combination of photodisintegration reactions, i.~e.\ $(\gamma,n)$,
$(\gamma,p)$ and $(\gamma, \alpha)$ reactions, in the explosive
scenario of the \emph{p} process \cite{Lamb92,Arno03}. An accurate
description of the nucleosynthesis within these processes demands
reliable input from nuclear physics.\\
Calculations of astrophysical reaction networks within the
\emph{p} process have to account for a huge number of reaction
rates. Many of the isotopes involved in the network are off the
valley of stability and, therefore, are not accessible by
experiments in many cases. But even for the stable nuclei
experimental data for photodisintegration reactions in the
astrophysically relevant energy region, i.~e.\ close to the
particle emission threshold, are rare. Hence, the reaction rates
mainly need to be adopted from theoretical predictions. It is
mandatory to test the predictive power of the calculations for
those isotopes, which can be studied in the laboratory, and to
prove the reliability of the predictions for the extrapolation to unstable nuclei.\\
The situation is much better for the \emph{s}~process. Since for
the unstable isotopes within the \emph{s}-process reaction network
the $\beta$-decay rate usually highly exceeds the neutron capture
rate, the reaction path follows the valley of stability and,
hence, mainly stable isotopes are involved. Therefore, extensive
experimental studies have been carried out in neutron capture
experiments in the last decades and neutron capture cross sections
for a wide range of isotopes have become available with
uncertainties of only a few percent \cite{Bao00,KADONIS}. With
this enormous amount of nuclear physics input the understanding of
 \emph{s}-process nucleosynthesis and its astrophysical sites
has largely improved. However, for some unstable isotopes along
the reaction path, the so-called \emph{branching points},
$\beta$-decay and neutron capture become competetive. Since the
branching ratio of $\beta$-decay and neutron capture is highly
sensitive to the physical conditions, e.~g.\ temperature and
neutron density, in the stellar environment during the \emph{s}
process, these branching points serve as an excellent test for
\emph{s}-process nucleosynthesis models. Thus, it is essential to
determine the neutron capture cross sections of these isotopes
with high accuracy. Unfortunately, due to their short half-life
neutron capture experiments can hardly be performed and, hence,
further improvements of nucleosynthesis models are still hampered
by the relatively large uncertainties of the cross section
predictions for these isotopes \cite{Kaep06}. Nevertheless,
information for the neutron capture cross sections of branching
points can be derived from studying the $(\gamma, n)$ reaction of
the stable neighbouring nucleus. The idea to determine
$(n,\gamma)$ reaction
rates via photodisintegration experiments has already been presented in \cite{Sonn03a}.\\
Experiments with real photons provide a well-suited tool to study
photodisintegration cross sections and ground-state reaction
rates. In previous experiments we have concentrated our
investigations on isotopes of mass $A\geq 186$ and confirmed the
validity of several theoretical predictions in this mass region
\cite{Vogt01b, Vogt02,Sonn03a,Sonn04b,Muel06}.
Average deviations between experiment and theory were typically less than 30\%.\\
In this paper we want to address the photodisintegration reactions
in the rare earth region (see Fig.~\ref{fig:process_flow}). This
region is of particular interest for the \emph{s} process, because
the isotopic abundance pattern is shaped by several branchings at
the unstable isotopes of Nd, Pm, Sm, Eu and Gd. Since the relative
abundances of the involved isotopes are known with an uncertainty
of better than two percent \cite{Ande89}, this mass region is
exceptionally suited to test the stellar nucleosynthesis models
with high accuracy. To provide precise nuclear physics input,
neutron capture cross sections of several branching points have
been measured with high accuracy in the last years
\cite{Jaag95,Reif03, Wiss06,Marr06}. But still one has to rely on
theoretical predictions of the cross sections of short-lived
branching points, e.~g.\ $^{147}$Nd and $^{153}$Gd. Therefore, we
have investigated the inverse photoneutron reactions of $^{148}$Nd
and $^{154}$Gd in order to
improve the reliability of these predictions.\\
Furthermore, the abundances of some isotopes in this region
receive large contributions from the \emph{p} process, e.~g.\ a
33\% contribution is predicted for $^{152}$Gd \cite{Ande89}. For a
comprehensive description of the nucleosynthesis in this mass
region these contributions cannot be neglected and need to be
studied in detail. Hence, the aim of this work is also to provide
reliable data for the photoneutron reaction rates of some selected
isotopes to prove the predictive
power of the theoretical calculations in this mass region.\\

We have performed photoactivation experiments to study the
$(\gamma,n)$ reactions of the isotopes $^{148,150}$Nd, $^{154}$Sm
and $^{154,160}$Gd in the astrophysically relevant energy region
close to the neutron emission threshold at the superconducting
Darmstadt electron linear accelerator S-DALINAC \cite{Rich96}. Our
experimental setup is discussed in detail in Sec.~\ref{sec:Setup}.
We explain the data analysis and the approach to derive
astrophysically relevant ground-state reaction rates directly from
the experiment by approximating a thermal Planck spectrum by a
superposition of different bremsstrahlung spectra in
Sec.~\ref{sec:DataAnalysis}. The results are presented in
Sec.~\ref{sec:Results} and compared to two theoretical
calculations based on the framework of the Hauser-Feshbach theory.

\section{Experimental setup}\label{sec:Setup}
\begin{figure}[htbp]
  \includegraphics[width=0.95\columnwidth]{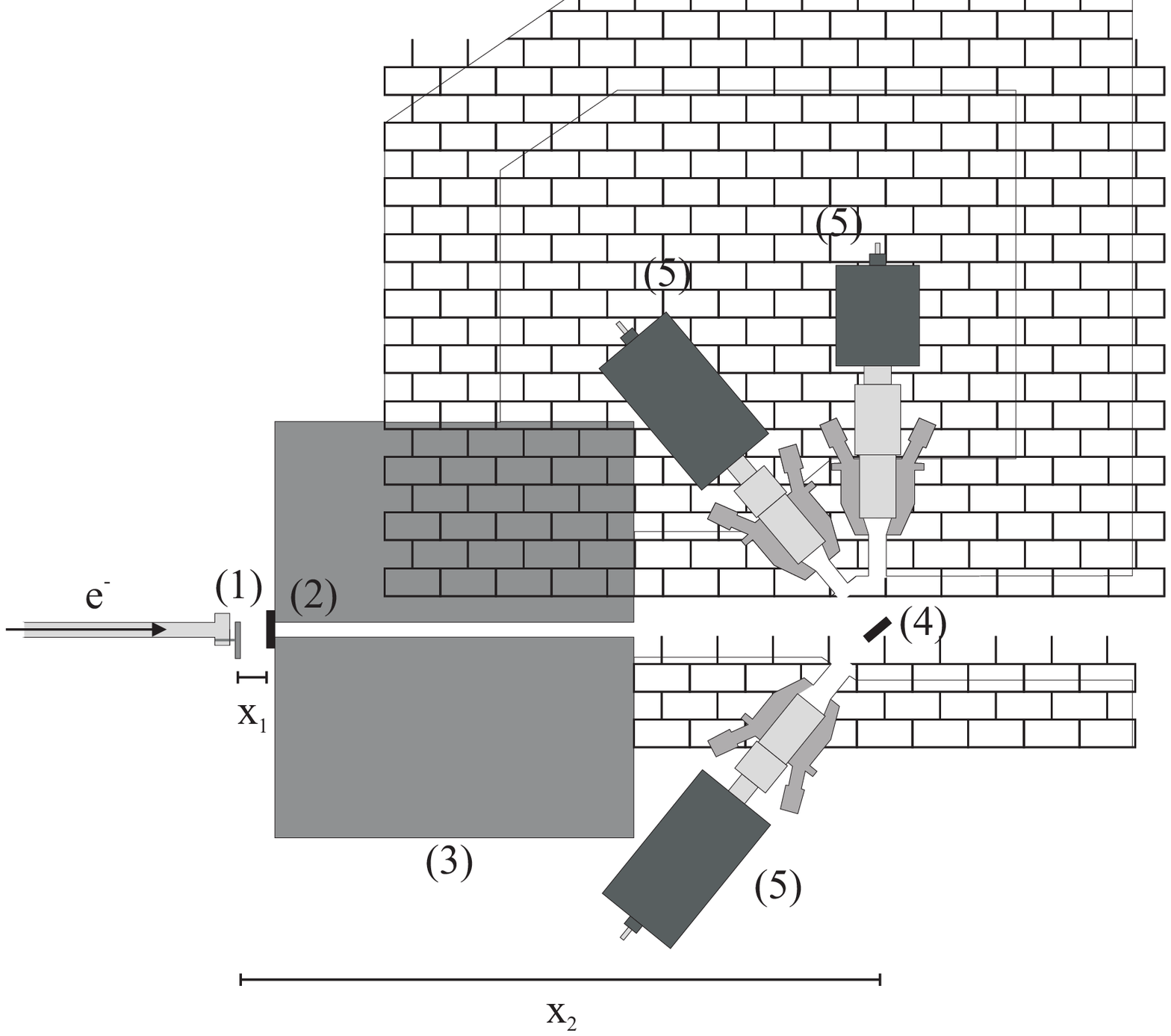}\\
  \caption{Schematic layout of the photoactivation setup. A monoenergetic electron beam produces bremsstrahlung in a massive copper radiator with a thickness of 1.2~cm (1).
  The photons activate the target of interest ($x_1\approx 5$~cm), which is sandwiched between several calibration targets (2). Behind a thick copper collimator system (3) a boron target (4)
  is mounted ($x_2\approx 150$~cm) to monitor the photon flux via the photon scattering reaction $^{11}$B$(\gamma,\gamma')$ with actively-shielded HPGe detectors (5).}\label{fig:Setup}
\end{figure}

We have irradiated naturally composed neodymium, gadolinium and
samarium targets at the superconducting electron linear
accelerator S-DALINAC. The experimental setup is illustrated in
Fig.~\ref{fig:Setup}. A monoenergetic electron beam of energy
$E_0$ is stopped completely in a thick copper radiator and
produces a continuous spectrum of bremsstrahlung photons with a
maximum energy of $E_\mathrm{max}=E_0$. The photons irradiate the
target, which is mounted directly behind the radiator. Thin
metallic discs ($m=40-50$~mg) were used as neodymium and
gadolinium targets, whereas samarium was available as pressed
pills of Sm$_2$O$_3$ powder ($m=500-1750$ mg). Each target had a
diameter of 20~mm. The target specifications are summarized in
Tab.~\ref{tab:Targets}.

\begingroup
\squeezetable
\begin{table*}[htbp]
  \centering
  \setlength{\extrarowheight}{2.pt}
\begin{tabular}{ccccccc}
\hline\hline Isotope \quad & \quad Form \quad & \quad Abundance
[\%] \quad    & \quad Weight [mg] \quad  & \quad $S_n$ [keV] \quad
& \quad
Analyzed $\gamma$-transition [keV]\quad & \quad $I_\gamma$ [\%]\\
\hline
   $^{148}$Nd \quad        & \quad metal foil \quad & \quad 5.76 \quad   & \quad 40-50 \quad     &  \quad 7332.9 \quad     & \quad 91.1  \quad  & \quad $27.9\pm 1.1$ \quad \\
   \vspace{0.3cm}
                            &                       &                    &                       &                         & \quad 531.0 \quad  & \quad $13.1\pm 0.9$ \quad \\
   $^{150}$Nd \quad        & \quad metal foil \quad & \quad 5.64 \quad   & \quad 40-50 \quad     &  \quad 7379.9 \quad     & \quad 114.3 \quad  & \quad $19.2\pm 1.5$ \quad \\
                            &                       &                    &                       &                         & \quad 155.9 \quad  & \quad \, $5.93\pm 0.31$ \quad \\
                            &                       &                    &                       &                         & \quad 211.3 \quad  & \quad $25.9\pm 1.4$ \quad \\
                            &                       &                    &                       &                         & \quad 267.7 \quad  & \quad \, $6.03\pm 0.28$ \quad \\
   \vspace{0.3cm}
                            &                       &                    &                       &                         & \quad 270.1 \quad  & \quad $10.6\pm 0.5$ \quad \\
   $^{154}$Gd \quad         & \quad metal foil \quad& \quad 2.18 \quad   & \quad 40-50 \quad     &  \quad 8894.8 \quad     & \quad 97.4  \quad  & \quad $29.0\pm 0.8$ \quad  \\
   \vspace{0.3cm}
                            &                       &                    &                       &                         & \quad 103.2 \quad  & \quad $21.1\pm 0.6$ \quad  \\
   $^{160}$Gd \quad         & \quad metal foil \quad & \quad 21.86 \quad & \quad 40-50 \quad     &  \quad 7451.4 \quad     & \quad 58.0  \quad  & \quad $2.49\pm 0.07$ \quad  \\
                            &                        &                   &                       &                         & \quad 226.0 \quad  & \quad $0.217\pm 0.002$ \quad  \\
                            &                       &                    &                       &                         & \quad 348.3 \quad  & \quad $0.239\pm 0.003$ \quad  \\
   \vspace{0.3cm}
                            &                       &                    &                       &                         & \quad 363.6 \quad  & \quad $11.8\pm 0.1$\,\,\, \quad  \\
   $^{154}$Sm \quad         & \quad oxide \quad     & \quad 22.70 \quad  & \quad 500-1750 \quad  &  \quad 7967.6 \quad     & \quad 69.7  \quad  & \quad $4.73\pm 0.04$ \quad  \\
                            &                       &                    &                       &                         & \quad 97.4  \quad  & \quad $0.772\pm 0.019$ \quad  \\
   \vspace{0.3cm}
                            &                       &                    &                       &                         & \quad 103.2 \quad  & \quad $29.3\pm 0.2$\,\,\, \quad  \\
  $^{187}$Re \quad          & \quad metal foil \quad& \quad 62.60 \quad  & \quad 320-340 \quad   &  \quad 7363.0 \quad  & \quad 122.6  \quad    & \quad $0.603\pm 0.003$ \quad  \\
                            &                       &                    &                       &                      & \quad 137.2  \quad    & \quad $9.47\pm 0.30$ \quad  \\

  \hline\hline
\end{tabular}
\caption{Specifications of targets and calibration targets used
for the activation experiments. The intensities per decay
$I_\gamma$ were taken from \cite{ENSDF2006}.}\label{tab:Targets}
\end{table*}
\endgroup

The photon flux intensity was determined by observing the photon
scattering reaction $^{11}$B$(\gamma,\gamma')$ behind a copper
collimator system with actively-shielded high-purity germanium
(HPGe) detectors. To compare the photon flux intensity at the
target position in front of the collimator and at the position of
the $^{11}$B$(\gamma,\gamma')$ target behind the collimator thin
metallic rhenium discs have been irradiated simultaneously at both
positions to normalize the photon flux intensity via the
photodisintegration reaction $^{187}$Re$(\gamma,n)$, which was
studied in \cite{Muel06}. A detailed discussion of the photon flux
calibration is given in the following section.

\begin{figure}[htbp]
  \includegraphics[width=1.\columnwidth]{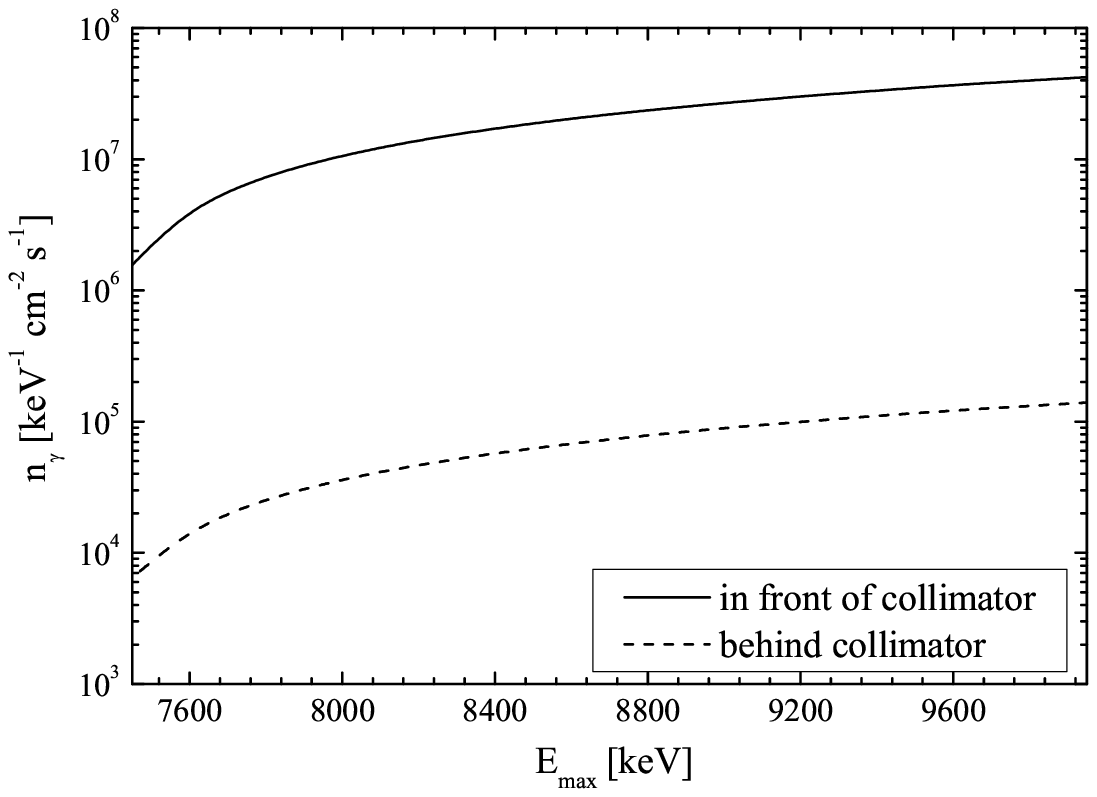}\\
  \caption{Available photon flux $n_\gamma$ at our photoactivation setup for a photon energy of {$E_\gamma=S_n(^{148}$Nd$)=7332.9$~keV}
  as a function of the maximum photon energy of the spectral distribution $E_\mathrm{max}$. A mean electron current of
  $I_e=25$~$\mu$A on the radiator has been assumed. The photon flux was obtained from a simulation using the Monte Carlo code {\sc Geant4} \cite{geant03} and confirmed
  experimentally by the two approaches discussed in the text.
  The photon flux in front of the collimator is about 270 times more intense than behind the collimator and the energy dependence is nearly equal.}\label{fig:FluxAbsolute}
\end{figure}

The advantage of irradiating the targets in front of the
collimator becomes clear from Fig.~\ref{fig:FluxAbsolute}. At this
position the much more intense photon flux guarantees a high
reaction yield for the irradiation even for small amounts of
target material and small photodisintegration cross sections.

The irradiation was performed at various energies $E_\mathrm{max}$
starting just above the neutron emission threshold of
$^{148,150}$Nd with $E_\mathrm{max}=7450$~keV. The energy was
increased in steps of 150 and 200~keV up to
$E_\mathrm{max}=9800$~keV. The
duration of each activation run was between 6 and 24 hours.\\

\subsection{Calibration of photon flux}\label{sec:PhotonFlux}
The spectral distribution of the photon flux is taken from a
simulation using the Monte Carlo code {\sc Geant4} \cite{geant03}.
The absolute normalization of the photon flux intensity can be
derived from the reaction yields of the photodisintegration
reaction $^{187}$Re$(\gamma,n)$ and the photon scattering reaction
$^{11}$B$(\gamma,\gamma')$, respectively.\\
The reaction yield $Y^{E_\mathrm{max}}_{(\gamma,n)}$ of the
photoactivation using bremsstrahlung with a maximum energy
$E_\mathrm{max}$ is given by:

\begin{equation}\label{eq:ReactionYield}
    Y_{(\gamma,n)}^{E_\mathrm{max}}=N_T\int{N_{\gamma}(E, E_\mathrm{max})\
    \sigma_{(\gamma,n)}(E)\ dE}\ ,
\end{equation}

where $N_T$ denotes the number of target nuclei, $N_\gamma(E,
E_\mathrm{max})$ the time-integrated photon flux
$\int_0^{t_\mathrm{act}}{n_\gamma(E, E_\mathrm{max}, t)}\ dt$ for
the duration of activation $t_\mathrm{act}$ and
$\sigma_{(\gamma,n)}$ is the photoneutron cross section. Using a
simulation $N_{\gamma,\mathrm{sim}}(E, E_\mathrm{max})$ for the
spectral distribution of the photon flux the normalization factor
$N_{\gamma,0}^{E_\mathrm{max}}$ for the simulation can then be
derived from:

\begin{equation}\label{eq:Normalization}
N_{\gamma,0}^{E_\mathrm{max}}=\frac{Y_{(\gamma,n)}^{E_\mathrm{max}}}{N_T\int{N_{\gamma,\mathrm{sim}}(E,
E_\mathrm{max})\
    \sigma_{(\gamma,n)}(E)\ dE}}\, .
\end{equation}

By irradiating $^{187}$Re both in front and behind the collimator
one can determine the normalization for the photon flux intensity
at both target positions from the well-known photoneutron cross
section of $^{187}$Re and the measured reaction yield
$Y_{(\gamma,n)}^{E_\mathrm{max}}$.

\begin{figure}[htbp]
  \includegraphics[width=1.\columnwidth]{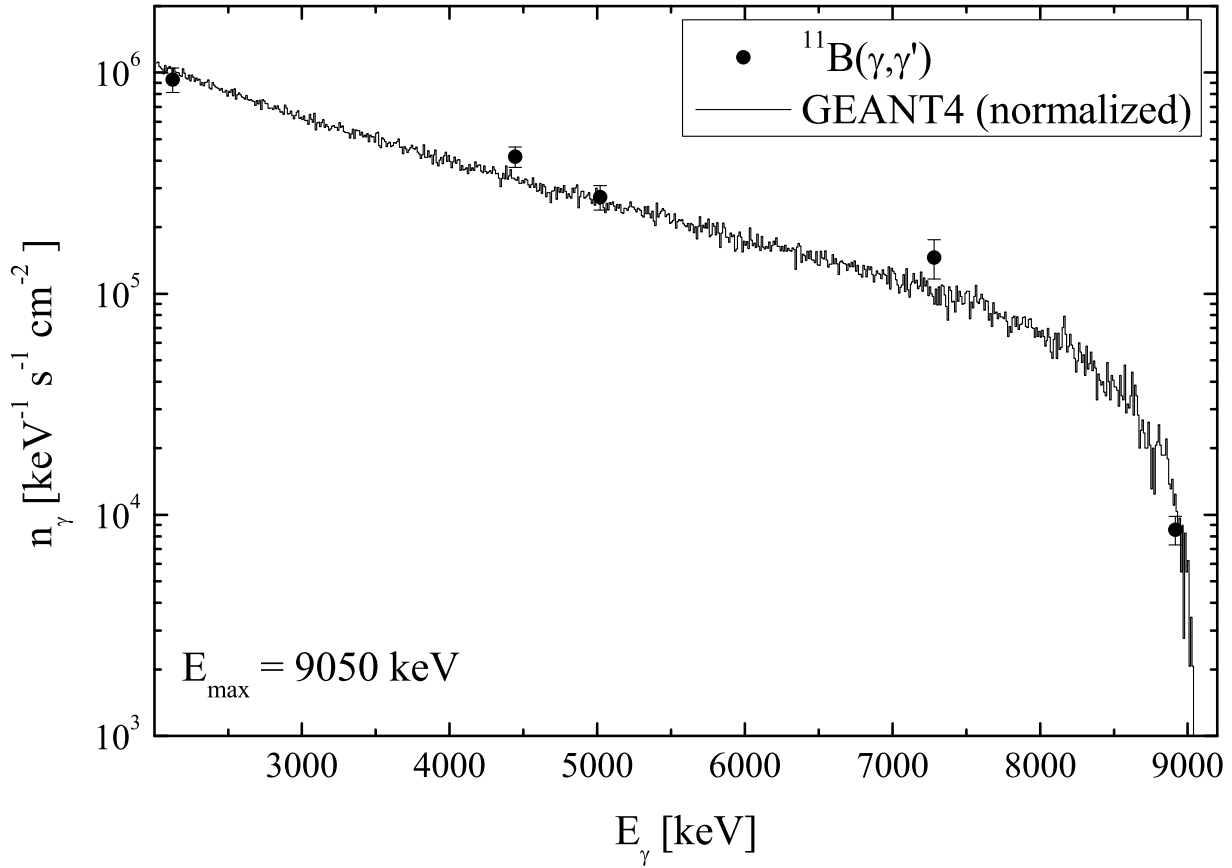}\\
  \caption{The simulated photon flux distribution at the target position behind the collimator for $E_\mathrm{max}=9050$~keV. The simulation was normalized
  to the experimental data obtained from the reaction yields of the photon scattering reaction $^{11}$B$(\gamma,\gamma')$ (see text for details).}\label{fig:FluxShape}
\end{figure}

A second approach for the normalization is to observe the photon
scattering reaction $^{11}$B$(\gamma,\gamma')$ with HPGe detectors
behind the collimator. The reaction yield for a certain transition
from a state of energy $E_i$ into a state of energy $E_j$ is given
by:

\begin{equation}\label{eq:PhotonScattering}
    Y_{i\rightarrow j}=N_{\mathrm{Bor}}\cdot
    N_{\gamma}(E_i,E_\mathrm{max})\cdot I_{i\rightarrow j}\ ,
\end{equation}

where $N_{\mathrm{Bor}}$ is the number of $^{11}$B nuclei in the
target and $I_{i\rightarrow j}$ denotes the integrated cross
section of the observed $\gamma$-transition. With the knowledge of
$I_{i\rightarrow j}$ one can directly determine the photon flux
intensity at various energies by observing different transitions.
These data points can then be used for the normalization of the
full photon spectrum. Figure \ref{fig:FluxShape} shows the
normalized photon flux distribution calculated with {\sc Geant4}
in comparison to the experimental data points of the
$^{11}$B$(\gamma,\gamma')$ reaction. A mean deviation of 10\% to
20\% between simulation and experimental data was found depending
on the maximum photon energy $E_\mathrm{max}$. In comparison with
former simulations using {\sc Geant3} \cite{GEANT} the shape of
the photon flux is correctly described by {\sc Geant4} and,
therefore, a correction procedure of the shape of the simulated
photon flux distribution close to $E_\mathrm{max}$ \cite{Vogt01b}
does not need to be applied anymore. This proves the reliability
of the simulation.

\begin{figure}[htbp]
  \includegraphics[width=1.\columnwidth]{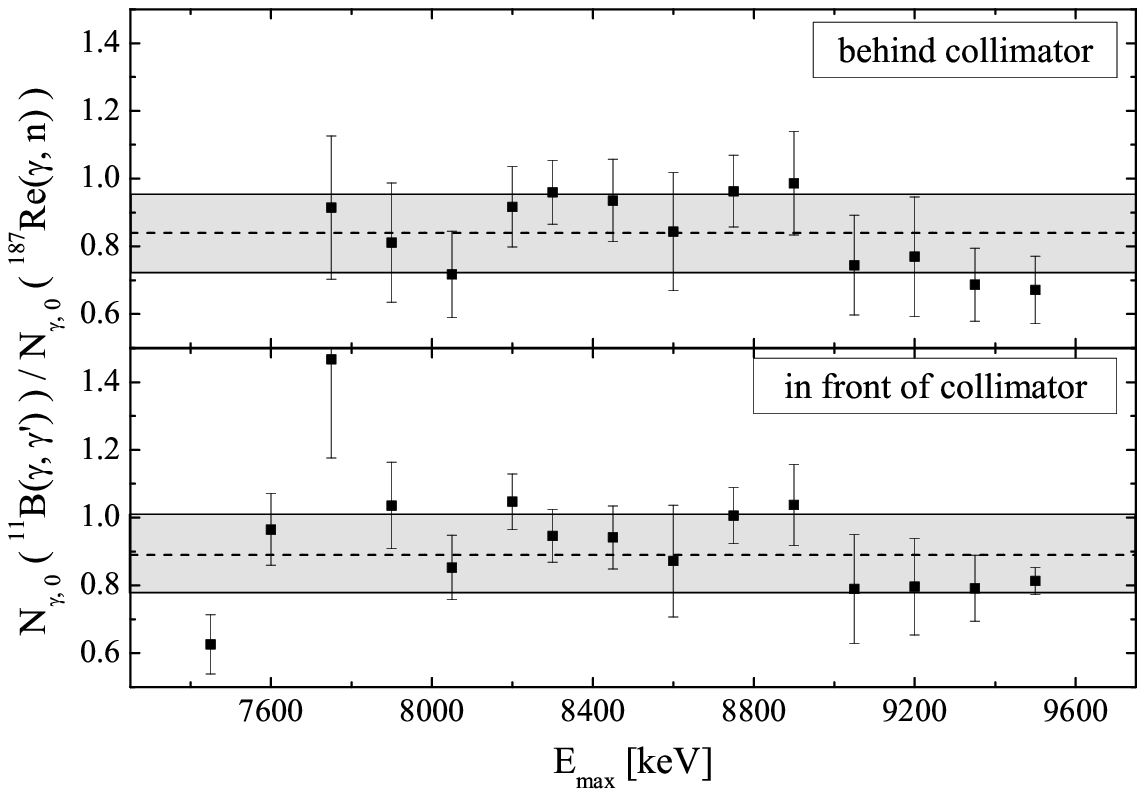}\\
  \caption{The ratio of the normalization factors $N_{\gamma,0}$ found for the photon flux intensities at different energies
  $E_\mathrm{max}$ using the reactions $^{187}$Re$(\gamma,n)$ and $^{11}$B$(\gamma,\gamma')$, respectively.
  Both reactions are in good agreement with regard to the experimental error bars and the uncertainty of the photoneutron cross section
  of $^{187}$Re (grey error band) \cite{Muel06}. The mean ratio of both approaches is indicated by the dashed line.
  At $E_\mathrm{max}=7450$~keV and $E_\mathrm{max}=7600$~keV the $^{187}$Re$(\gamma,n)$ reaction yield was too low to obtain data points behind the collimator.}\label{fig:FluxCalibration}
\end{figure}

Nevertheless, this approach only yields the normalization of the
photon flux intensity at the target position behind the
collimator, but we assumed that the same normalization factor was
valid at the target position in front of the collimator. In
addition, to confirm the normalization based on the
$^{11}$B$(\gamma,\gamma')$ reaction the photon flux intensity was
derived from the photodisintegration reaction
$^{187}$Re$(\gamma,n)$ at both target positions. As seen from
Fig.~\ref{fig:FluxCalibration} both approaches are in excellent
agreement.

\subsection{Determination of reaction yield}\label{sec:Reaction_Yield_Determination}
\begin{figure}[htbp]
  \includegraphics[width=1.\columnwidth]{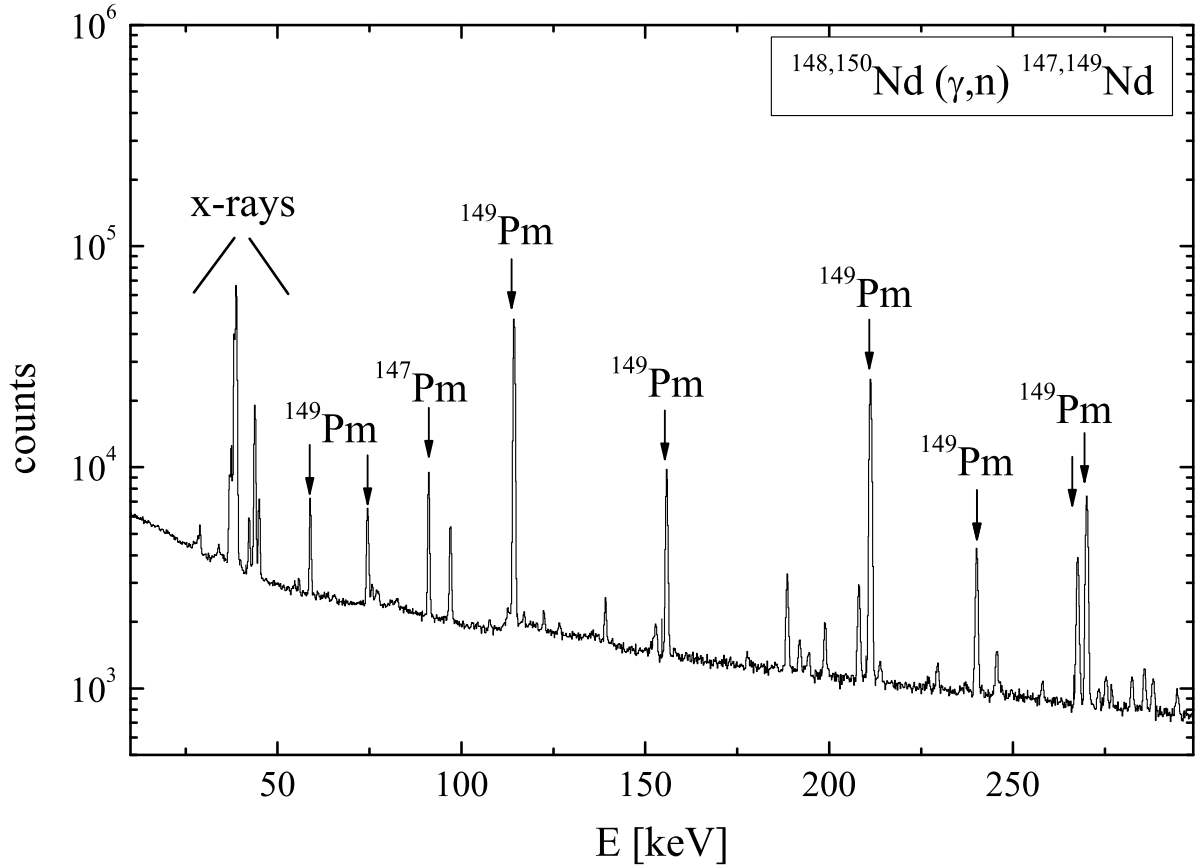}\\
  \caption{Typical $\gamma$-decay spectrum of a neodymium target irradiated at $E_\mathrm{max}=8600$~keV for a duration of about 13 hours.
  The spectrum was accumulated over a period of 3~hours at the LEPS setup. The strongest $\gamma$-transitions following the $\beta$-decays of
  $^{147}$Nd into $^{147}$Pm and $^{149}$Nd into $^{149}$Pm, respectively, are indicated.} \label{fig:Spectrum}
\end{figure}

In order to determine the reaction yield $Y_{(\gamma,n)}$, i.~e.\
the number of $(\gamma,n)$-reactions occurring during the
irradiation, the $\gamma$-transitions following the $\beta$-decays
of the produced unstable isotopes were measured offline with HPGe
detectors after the activation. The detectors were covered by
thick lead shielding to reduce natural background. The reaction
yield is directly proportional to the peak areas of the
corresponding $\gamma$-transitions in the spectra. The factor of
proportionality can be determined with regard to the activation
time, the detection efficiency and the $\gamma$-intensities. This
has been
discussed in detail in \cite{Vogt01b}.\\
Two different setups have been used to measure the activity of the
produced unstable isotopes. One setup consisted of two low-energy
photon spectrometers (LEPS), which are highly sensitive to
low-energy photons down to a few keV due to a very thin beryllium
entrance window. They were positioned face-to-face with a distance
of only 10 mm to each other. The targets were mounted directly
between the two detectors to obtain a high detection efficiency.
Figure \ref{fig:Spectrum} shows a spectrum of an irradiated
naturally composed neodymium target. Due to the excellent energy
resolution of the detectors ($\Delta
E^{\mathrm{FWHM}}_\gamma\approx0.5$~keV at $E_\gamma=100$~keV) the
observed $\gamma$-transitions could be clearly assigned to the
corresponding $\beta$-decays. However, due to the high detection
efficiency, summing effects of coincident $\gamma$-rays stemming
from the same decay cascade cannot be neglected and have to be
carefully corrected in the analysis of the reaction yield. For the
isotopes studied in this work a maximum correction of about 10\%
was found for the summing of $\gamma$-rays and x-rays in the case
of the electron capture of $^{153}$Gd.

\begin{figure}[htbp]
  \includegraphics[width=1.\columnwidth]{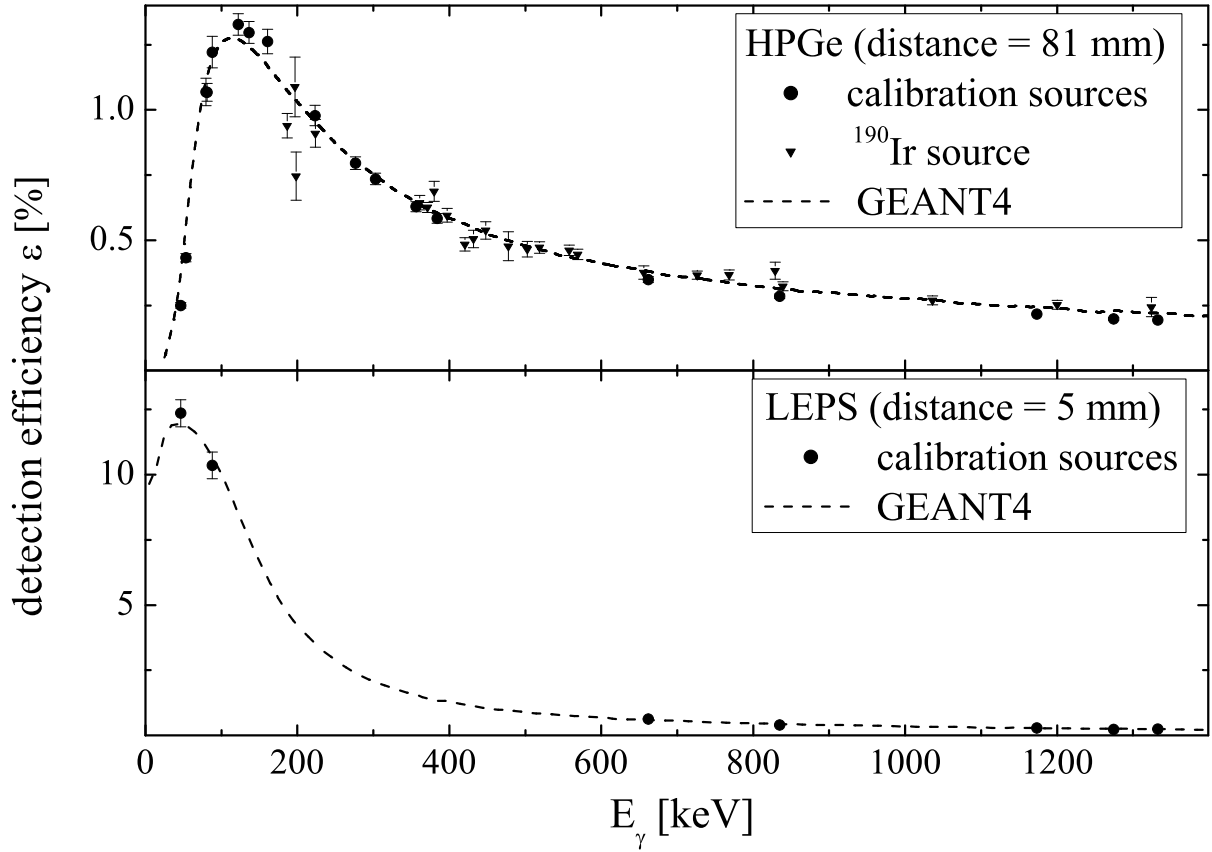}\\
  \caption{Absolute detection efficiency of the HPGe and LEPS setup. Standard calibration sources were used to normalize
  the efficiency curve calculated with {\sc Geant4}. Due to large summing effects of coincident $\gamma$-rays at
  small distances between detector and source, only those calibration sources without $\gamma$-cascades were used for
the normalization in case of the LEPS
setup.}\label{fig:Efficiency_Detectors}
\end{figure}

At the second setup a HPGe detector with 30\% efficiency relative
to a $3" \times 3"$ NaI detector was used. Here the targets were
mounted at a larger distance of 81~mm in front of the detector.
Thus, almost no summing effects occurred in the spectra. However,
due to the much smaller detection efficiency this second setup was
only used to verify the results
obtained at the LEPS setup for a few selected targets.\\
To determine the absolute efficiency of our setup, the efficiency
of the detectors was simulated in detail using {\sc Geant4}. In
addition, the efficiencies at certain energies between $14$~keV
and $1350$~keV were measured using standard calibration sources,
which served as normalization for the calculated efficiencies.
Moreover, a non-calibrated $^{190}$Ir source was used to confirm
the predicted energy dependence of the efficiency. Figure
\ref{fig:Efficiency_Detectors} shows that the normalized
simulations are in good agreement with the experimental data
points. From the uncertainties of the experimental data points and
the deviation of the normalized simulation to these data points,
it was estimated that the energy-dependent detection efficiency
can be determined with an uncertainty of better than 7\% for
energies of up to about 1500~keV.\\
The simulations were also used to account for the self-absorption
of the low-energy decay $\gamma$-rays within the target. Whereas
the corrections were almost negligible for the very thin neodymium
and gadolinium foils, corrections of up to 80\% had to be applied
for the relatively thick Sm$_2$O$_3$ targets. To test the
reliability of the simulations the self-absorption was also
measured at different photon energies for a variety of targets.
Deviations between measurement and simulation were found to be of
the order of a few percent.

\section{Data Analysis}\label{sec:DataAnalysis}

\subsection{Cross section determination}\label{sec:Cross-Section}
The reaction yield $Y_{(\gamma,n)}^{E_\mathrm{max}}$ is directly
proportional to the energy-integrated cross section $I_{\sigma}$,
which is given by the integral from Eq.~(\ref{eq:ReactionYield}).
Since bremsstrahlung is characterized by a continuous spectral
distribution, a deconvolution of the determined integrated cross
section $I_\sigma$ is not possible in general. Therefore, the
cross section $\sigma_{(\gamma,n)}$ cannot be derived directly.
However, if a theoretical prediction for $\sigma_{(\gamma,n)}$ is
adopted, one can calculate $I_\sigma$ from
Eq.~(\ref{eq:ReactionYield}) and then derive a normalization
factor $f$ for the prediction from a comparison to the
experimentally determined yields:

\begin{eqnarray}\label{eq:Cross-Section-Theo}
    f(E_{\mathrm{max}}) & = & \frac{I_\sigma^{\mathrm{Exp}}}{\int^{E_{\mathrm{max}}}_{S_n} N_\gamma(E,E_{\mathrm{max}})\cdot \sigma_{(\gamma,n)}^{\mathrm{Theory}}(E)\,
    dE}\,\, .
\end{eqnarray}

\begin{figure}[htbp]
  \includegraphics[width=1.\columnwidth]{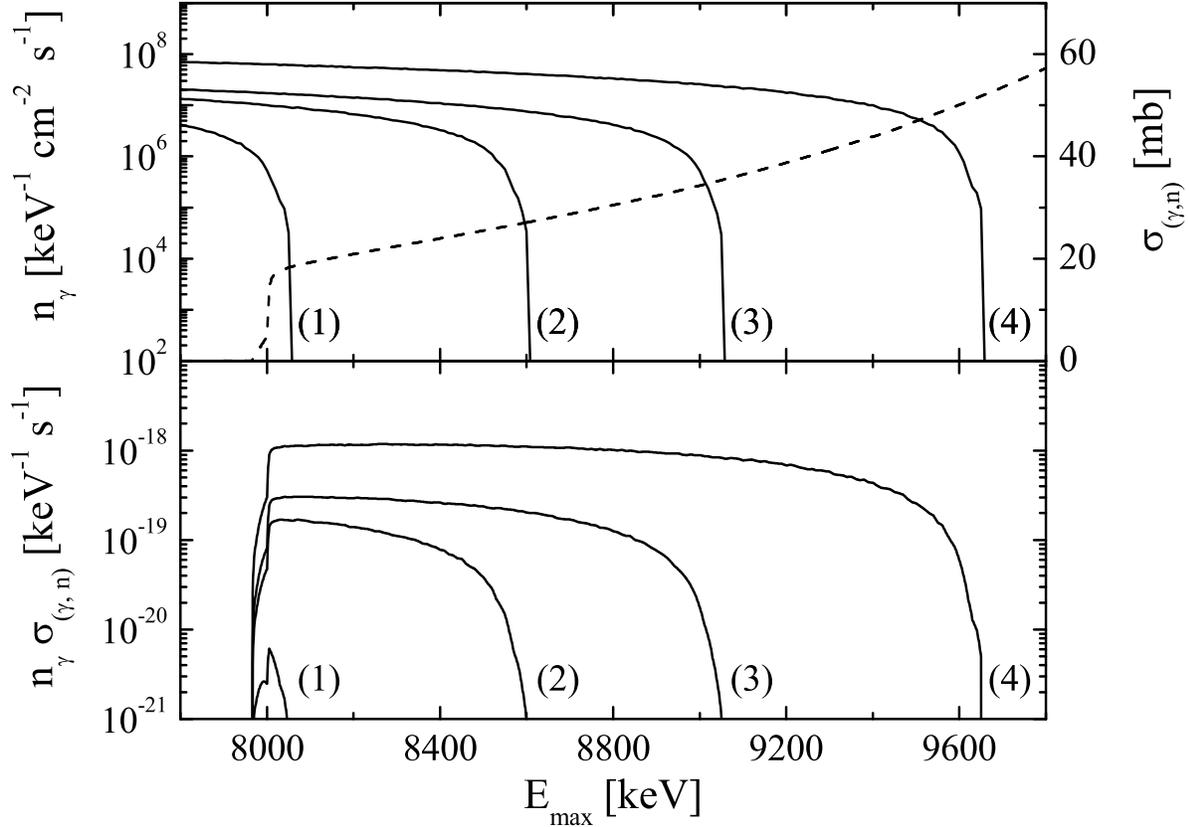}\\
  \caption{The top panel shows the simulated spectral distribution of bremsstrahlung at different energies $E_{\mathrm{max}}$ of (1)~8050~keV, (2)~8600~keV, (3)~9050~keV and
  (4)~9650~keV (solid lines) in comparison to the photoneutron cross section $\sigma_{(\gamma,n)}$ of $^{154}$Sm (dashed line) predicted by the NON-SMOKER$^{\mathrm{WEB}}$ code \cite{NONSMOKERWEB}.
  The product of $\sigma_{(\gamma,n)}$ and $n_\gamma$ yields the integrand of Eq.~(\ref{eq:ReactionYield}), which covers a broad energy range
close to the neutron separation energy as shown in the bottom
panel.}
  \label{fig:Integrated-Cross-Section-Integrand}
\end{figure}

By irradiating at different energies $E_{\mathrm{max}}$ and
deriving $f$ for each energy one can test the theoretical
prediction for $\sigma_{(\gamma,n)}$ within different energy
ranges. If $\sigma_{(\gamma,n)}$ is accurately described, then $f$
should be independent of $E_{\mathrm{max}}$ and close to unity.
However, our photoactivation experiments using bremsstrahlung only
have limited sensitivity to the shape of $\sigma_{(\gamma,n)}$,
since the normalization is an average over a wide energy range as
illustrated in Fig.~\ref{fig:Integrated-Cross-Section-Integrand}.

\subsection{Ground-state reaction rates}\label{sec:Reaction-Rates}
The photodisintegration reaction rate $\lambda$ for a nucleus in a
thermal photon bath is given by

\begin{equation}\label{eq:Reaction-Rate}
    \lambda(T)=\int_0^{\infty} n_\gamma^{\mathrm{Planck}}(E,T)\,
    \sigma_{(\gamma,n)}(E)\,dE,
\end{equation}

where $\sigma_{(\gamma,n)}$ is the photoneutron cross section and
$n_\gamma^{\mathrm{Planck}}$ the photon flux per energy interval
given by the Planck distribution

\begin{equation}\label{eq:Planck}
n_\gamma^{\mathrm{Planck}}(E,T) = c\, (\frac{1}{\pi})^2
(\frac{1}{\hbar c})^3 \frac{E^2}{\exp{E/kT}-1}.
\end{equation}

Although it is not possible to produce a thermal photon bath at
\emph{p}-process conditions, i.~e.\ with the intensities resulting
from temperatures between 2 and $3\times 10^9$~K, and thus to
measure stellar reaction rates in the laboratory, we can use two
different approaches to determine the ground-state reaction rates
$\lambda^{\mathrm{g.s.}}$ in the experiment. The obvious way is to
use the normalized cross section $\sigma_{(\gamma,n)}=f\cdot
\sigma_{(\gamma,n)}^\mathrm{Theory}$ from the analysis described
in the preceding section and then calculate the integral of
Eq.~(\ref{eq:Reaction-Rate}). However, using this method one has
to rely on the adopted theoretical prediction of the energy
dependence of the cross section. This might lead to large
systematic uncertainties, if the predicted shape of the cross
section deviates significantly from its real shape, e.~g.\ if some
resonances above the neutron emission threshold are omitted in the
theoretical description. Hence, an approach is preferred where the
reaction rates can be directly determined from the experimental
data without the need of any theoretical input. This can be
achieved by approximating the Planck spectrum at temperature $T$
with a superposition of several bremsstrahlung spectra
$n_\gamma^{\mathrm{Brems}}(E,E_{\mathrm{max}}^i)$ at different
energies $E_{\mathrm{max}}$ \cite{Mohr00b}

\begin{equation}\label{eq:Superposition}
    n_\gamma^{\mathrm{Planck}}(E,T)\approx \sum_i
    a_i(T)\, n_\gamma^{\mathrm{Brems}}(E,E_{\mathrm{max}}^i),
\end{equation}

where $a_i(T)$ are temperature-dependent weighting coefficients.
With this approximation Eq.~(\ref{eq:Reaction-Rate}) can then be
written as

\begin{eqnarray}\label{eq:Approx-Reaction-Rate}
   \lambda(T)&\approx&\sum_i a_i(T)\int
    n_\gamma^{\mathrm{Brems}}(E,E_{\mathrm{max}}^i)
    \sigma_{(\gamma,n)}(E)dE \\
   &=&\sum_i a_i(T)\, I_{\sigma,i}^\mathrm{Exp}.
\end{eqnarray}

\begin{figure}[htbp]
  \includegraphics[width=1.\columnwidth]{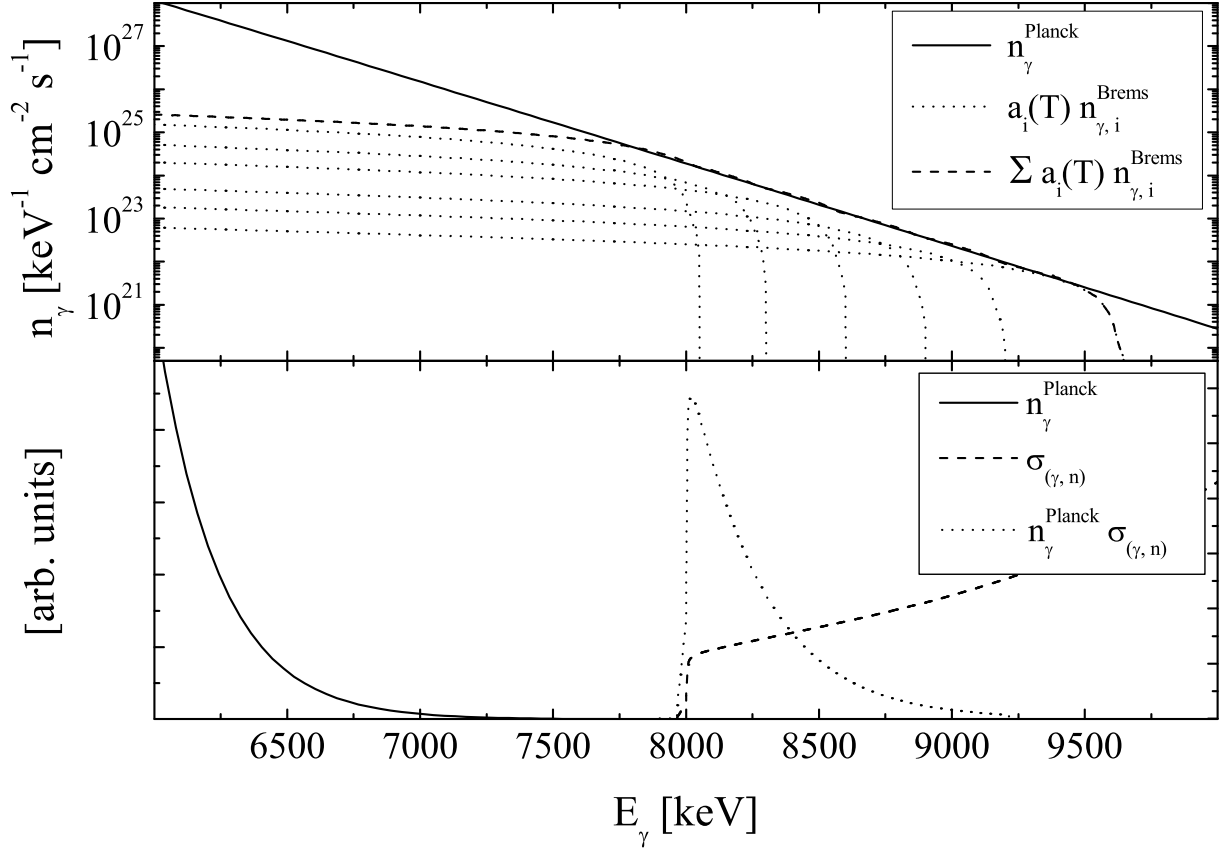}\\
  \caption{The top panel shows a thermal Planck spectrum at $T=2.5\times 10^9$~K. The weighted
  sum of different bremsstrahlung spectra yields a good approximation of the Planck spectrum within
  the Gamow-like window. The bottom panel shows that this energy region is defined by the product
  of photon flux $n_\gamma^{\mathrm{Planck}}$ and photoneutron cross section $\sigma_{(\gamma,n)}$ \cite{Mohr00b}. In this figure $\sigma_{(\gamma,n)}$
  was calculated with the NON-SMOKER$^{\mathrm{WEB}}$ code \cite{NONSMOKERWEB} for the case of $^{154}$Sm.}\label{fig:Planck_Approximation}
\end{figure}

Since the integrated cross sections $I_{\sigma,i}^\mathrm{Exp}$
are directly determined from the experiment the reaction rates can
be obtained without further assumptions on the energy dependence
of the cross section. Therefore, this analysis is free of
systematic uncertainties stemming from the uncertainties of any
cross section prediction. The deviation between the approximated
and the real Planck spectrum in the relevant energy region for
astrophysical studies close to the neutron threshold energy (the
so-called \emph{Gamow-like window} \cite{Mohr00b}) is of the order
of 10\%, depending on how many bremsstrahlung spectra are used for
the approximation (see Fig.~\ref{fig:Planck_Approximation}).

\section{Results}\label{sec:Results}

\subsection{Normalization of theoretical
predictions}\label{sec:Normalization} 
\begin{figure*}[htbp]
  \includegraphics[width=0.95\textwidth]{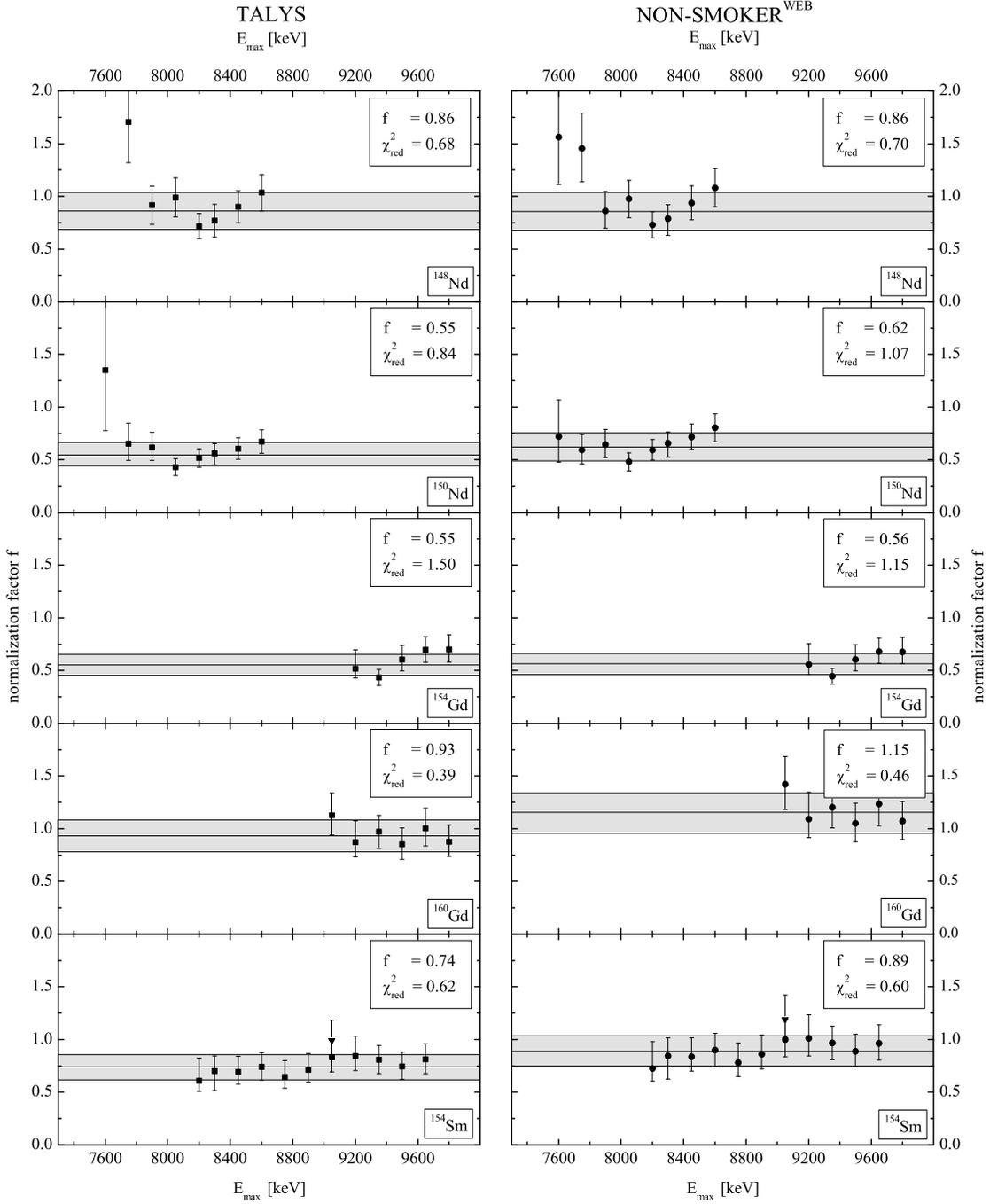}\\
  \caption{Normalization factors $f$ derived for two different theoretical predictions of the photoneutron cross section at various energies $E_\mathrm{max}$. The
  grey error band denotes the experimental uncertainty $\Delta f_\mathrm{total}$ discussed in the text. The triangular data points at $E_\mathrm{max}=9050$~keV
  for $^{154}$Sm are independent remeasurements to test the reproducibility of our experimental results.}\label{fig:WQ_Normalization}
\end{figure*}

\begin{figure*}[htbp]
  \includegraphics[width=1.\textwidth]{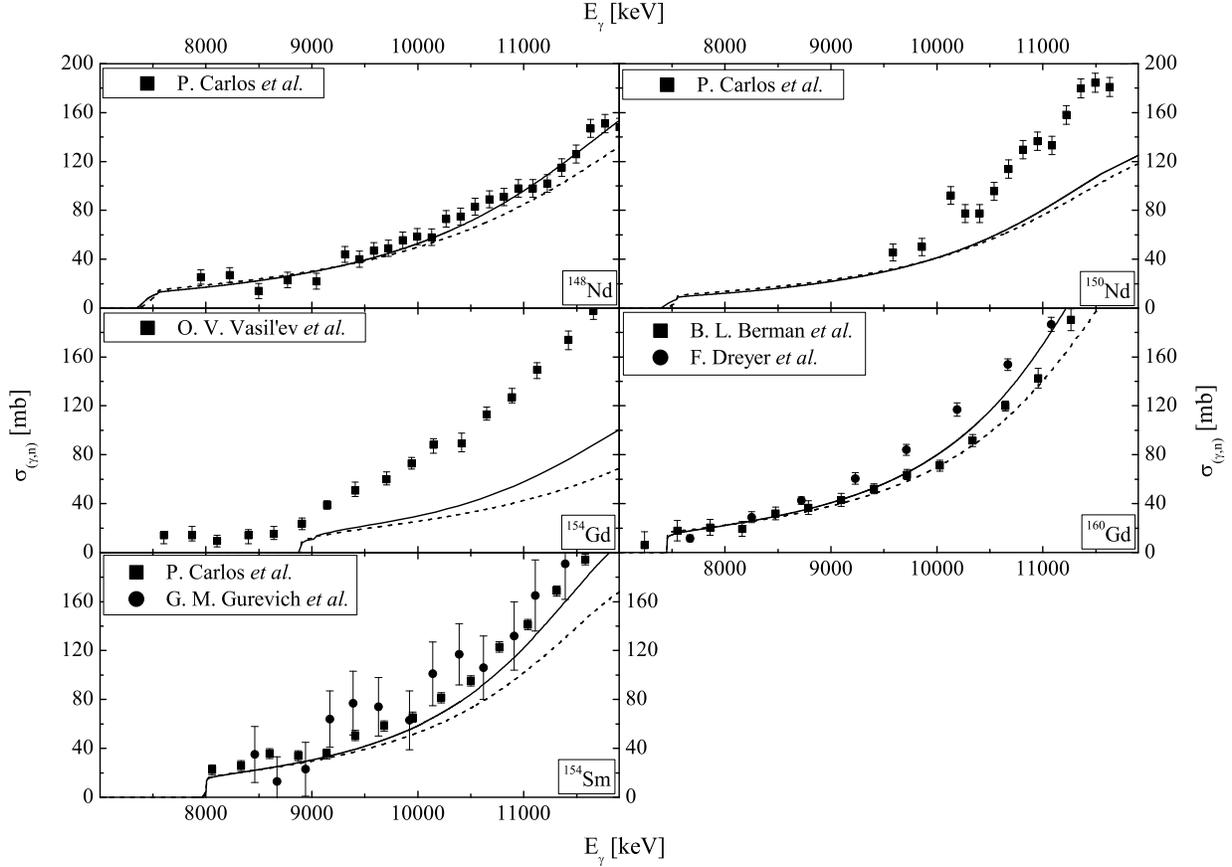}\\
  \caption{Comparison of experimental data for the photoneutron cross section from various photodisintegration experiments
  \cite{Carl71,Carl74,Vasi71,Gure81,Drey72,Berm69} and theoretical predictions using the TALYS (dashed line) and
  NON-SMOKER$^{\mathrm{WEB}}$ code (solid line). The theoretical calculations were normalized with the factors presented
  in Tab.~\ref{tab:Normalization_Results}. Note that the error bands of the
  theoretical predictions due to the experimental uncertainties of the
  applied normalization factors have been omitted in the graphs.}\label{fig:WQ_Experiment}
\end{figure*}

We have determined
normalization factors for two different theoretical predictions of
the photoneutron cross section $\sigma_{(\gamma,n)}$, which were
calculated with the NON-SMOKER$^{\mathrm{WEB}}$ code by Rauscher
\cite{NONSMOKERWEB} and the TALYS code by Koning \emph{et al.}\
\cite{TALYS}. Both theoretical predictions are based on the
Hauser-Feshbach formalism presuming that the nuclear level density
in the energy range close to the neutron threshold is high enough
for a statistical treatment. Different results between the two
calculations can mainly be ascribed to the nuclear physics input
used in the codes, e.~g.\ the neutron optical potential, nuclear
level densities and the $\gamma$-ray
strength function.\\
To compare the results of both predictions we used the default
input parameters recommended by the authors of the codes. These
input parameters are mainly intended to provide a global
description for a wide range of isotopes. The
NON-SMOKER$^{\mathrm{WEB}}$ code involves the neutron optical
potential published by Jeukenne \emph{et al.}\ \cite{Jeuk77} with
a low-energy modification by Lejeune \cite{Lejeune80}. The
$\gamma$-ray strength function is based on a description of
Thielemann and Arnould \cite{Thie83} using experimental GDR
energies and widths if available  and the low-energy modification
of the GDR Lorentzian by McCullagh \emph{et al.}\ \cite{McCu81}.
For the nuclear level density a global parametrization within the
back-shifted Fermi-gas formalism by Rauscher \emph{et al.}\
\cite{Raus97} is applied. The TALYS code uses the neutron optical
model potential parameterizations of Koning and Delaroche
\cite{Koning03}. The $\gamma$-ray strength function is obtained
from the compilation by Kopecky and Uhl \cite{Kope90} and the
nuclear level density is also based on an approach using the
Fermi-gas model \cite{Ericson60}. Both the TALYS code and the
NON-SMOKER$^{\mathrm{WEB}}$ code employ the Constant Temperature
Model from Gilbert and Cameron \cite{Gilb65} for the nuclear level
density to avoid the divergence of the Fermi-gas model
at low excitation energies.\\
The experimentally determined normalization factors $f$ of these
calculations for different energies $E_\mathrm{max}$ are presented
in Fig.~\ref{fig:WQ_Normalization} and the results are summarized
in Tab.~\ref{tab:Normalization_Results}. Both statistical model
codes are in fair agreement with our experimental data points. A
$\chi^2$-test confirmed that the normalization factors derived for
each isotope can be assumed to be independent of the energy
$E_\mathrm{max}$ within the experimental uncertainties. For the
absolute values of the cross sections we found a mean deviation of
24\% and 27\% between theory and experiment and a mean
normalization factor $f$ of 0.82 and 0.73 for the
NON-SMOKER$^{\mathrm{WEB}}$ and the TALYS calculation,
respectively. Thus, the predictions seem to slightly overestimate
the photoneutron cross sections in the rare earth region.\\
The quoted uncertainties $\Delta f_{\mathrm{sys}}$ and $\Delta
f_{\mathrm{fit}}$ denote the systematic and statistical
uncertainties of the measurement, respectively. They are discussed
in more detail in Sec.~\ref{sec:Error_discussion}. $\Delta
f_{\mathrm{fit}}$ was derived from averaging the normalization
factors $f$ measured at different energies $E_\mathrm{max}$ with
regard to the statistical uncertainties of each individual data
point. The total uncertainty $\Delta f_{\mathrm{total}}$ is
calculated by a Gaussian error propagation of $\Delta
f_{\mathrm{sys}}$ and $\Delta f_{\mathrm{fit}}$. Although the
experimental uncertainty is about 20\%, it has to be pointed out
that most of the isotopes were simultaneously irradiated. Hence,
systematic uncertainties are significantly reduced when comparing
the normalization factors of these isotopes relative to each
other. Therefore, the discrepancy between
experimental data and theory cannot only be due to experimental uncertainties.\\
Figure \ref{fig:WQ_Experiment} shows results from various
experiments for the photoneutron cross section in the energy
region of the giant dipole resonance. These data have been
obtained by using several experimental techniques like absorption
and activation measurements using bremsstrahlung, but also direct
measurements of the energy dependence of the cross section using a
quasi-monoenergetic photon beam produced by the annihilation in
flight of monoenergetic positrons. Details of these experiments
are given in \cite{Carl71,Carl74,Vasi71,Gure81,Drey72,Berm69}. The
theoretical predictions for the photoneutron cross sections
normalized with the factors found in our experiment have been
compared to these data. For the isotopes $^{148}$Nd, $^{154}$Sm,
and $^{160}$Gd the normalized calculations appear to be slightly
below the experimental data points, but are still fully consistent
with these data points within the quoted uncertainties of the
derived normalization factors. Furthermore, larger deviations have
been found for $^{150}$Nd and $^{154}$Gd. Unfortunately, in the
case of $^{150}$Nd a comparison close to the neutron emission
threshold is not possible, since no experimental data is available
in this energy region. For $^{154}$Gd a non-negligible
photoneutron cross section was even stated below the neutron
separation energy of $S_n=8894.8$~keV in \cite{Vasi71}. This
indicates systematic uncertainties of these experimental data and
might explain the discrepancy to the normalized predictions of
this work.

\begin{table*}[htbp]
  \centering
  \setlength{\extrarowheight}{2.pt}
  \begin{tabular}{ccccccccc}
  \multicolumn{1}{c}{}    & \multicolumn{4}{c}{TALYS}     & \multicolumn{4}{c}{NON-SMOKER$^{\mathrm{WEB}}$}\\
  \hline\hline
  \quad Isotope        \quad \quad & \quad \quad   $f$ \quad & \quad   $\Delta f_\mathrm{total}$   \quad & \quad  $\Delta f_\mathrm{fit}$  \quad & \quad   $\Delta f_\mathrm{sys}$  \quad \quad & \quad \quad $f$   \quad & \quad   $\Delta f_\mathrm{total}$   \quad & \quad  $\Delta f_\mathrm{fit}$  \quad & \quad  $\Delta  f_\mathrm{sys}$ \quad \quad\\
  \hline\hline
  \quad $^{148}$Nd    \quad \quad & \quad \quad  0.86 \quad & \quad   0.176                       \quad & \quad  0.052                    \quad & \quad   0.169                    \quad \quad & \quad \quad 0.86 \quad  & \quad   0.176                       \quad & \quad  0.052                    \quad & \quad  0.169 \quad \quad\\
  \quad $^{150}$Nd    \quad \quad & \quad \quad  0.55 \quad & \quad   0.111                       \quad & \quad  0.037                    \quad & \quad   0.105                    \quad \quad & \quad \quad 0.62 \quad  & \quad   0.127                       \quad & \quad  0.047                    \quad & \quad  0.118 \quad \quad\\
  \quad $^{154}$Gd    \quad \quad & \quad \quad  0.55 \quad & \quad   0.102                       \quad & \quad  0.052                    \quad & \quad   0.088                    \quad \quad & \quad \quad 0.56 \quad  & \quad   0.100                       \quad & \quad  0.045                    \quad & \quad  0.090 \quad \quad\\
  \quad $^{160}$Gd    \quad \quad & \quad \quad  0.93 \quad & \quad   0.151                       \quad & \quad  0.039                    \quad & \quad   0.145                    \quad \quad & \quad \quad 1.15 \quad  & \quad   0.186                       \quad & \quad  0.051                    \quad & \quad  0.179 \quad \quad\\
  \quad $^{154}$Sm    \quad \quad & \quad \quad  0.74 \quad & \quad   0.119                       \quad & \quad  0.029                    \quad & \quad   0.115                    \quad \quad & \quad \quad 0.89 \quad  & \quad   0.143                       \quad & \quad  0.034                    \quad & \quad  0.139 \quad \quad\\
  \hline\hline
  \end{tabular}

  \caption{Normalization factors derived from the experimental data for the theoretical predictions of the photoneutron cross section using the
  TALYS and NON-SMOKER$^{\mathrm{WEB}}$ code. The uncertainties $\Delta f_\mathrm{fit}$ and $\Delta f_\mathrm{sys}$ are discussed in detail in the text.
  $\Delta f_\mathrm{total}$ denotes the total experimental uncertainty derived from a Gaussian error propagation of $\Delta f_\mathrm{fit}$ and $\Delta f_\mathrm{sys}$.}\label{tab:Normalization_Results}

\end{table*}

\begin{table*}[h!tbp]
  \centering
  \setlength{\extrarowheight}{4.pt}
  \begin{tabular}{lccccc}
    \hline\hline
                            & \quad $^{148}$Nd \quad    & \quad $^{150}$Nd \quad    & \quad $^{154}$Gd \quad    & \quad $^{154}$Sm \quad    \\
    \hline
    $T=2.0\times 10^9$~K    & & & & &\\
    Upper energy limit  & \quad 8380 \quad   & \quad 8440 \quad    & \quad 9850 \quad    &  \quad 8930 \quad    \\
    \hline
  $\lambda^{\mathrm{g.s.}}_\mathrm{Exp}$                & \quad $5.97\cdot 10^{-3}$ \quad   & \quad $3.52\cdot 10^{-3}$ \quad  & \quad $2.74\cdot 10^{-6}$ \quad    & \quad $4.78\cdot 10^{-4}$ \quad \\
  $\Delta\lambda^{\mathrm{g.s.}}_\mathrm{Exp,Yield}$    & \quad $1.10\cdot 10^{-3}$ \quad  & \quad $0.64\cdot 10^{-3}$ \quad  & \quad $0.50\cdot 10^{-6}$    \quad  & \quad $0.85\cdot 10^{-4}$    \quad \\
  $\Delta\lambda^{\mathrm{g.s.}}_\mathrm{Exp,Approx}$   & \quad $1.64\cdot 10^{-3}$ \quad  & \quad $0.85\cdot 10^{-3}$ \quad  & \quad $0.48\cdot 10^{-6}$    \quad  & \quad $0.54\cdot 10^{-4}$    \quad \\
  $\lambda^{\mathrm{g.s.}}_\mathrm{TALYS}$   & \quad $1.09\cdot 10^{-2}$ \quad  & \quad $9.10\cdot 10^{-3}$ \quad  & \quad  $5.10\cdot 10^{-6}$    \quad            & \quad $7.87\cdot 10^{-4}$    \quad \\
  $\lambda^{\mathrm{g.s.}}_\mathrm{N.S.}$   & \quad $1.20\cdot 10^{-2}$ \quad  & \quad $9.44\cdot 10^{-3}$ \quad  & \quad  $5.01\cdot 10^{-6}$    \quad &  \quad $6.53\cdot 10^{-4}$    \quad \\
  \hline\hline
      $T=2.5\times 10^9$~K & & & & &\\
      Upper energy limit  & \quad 8655 \quad   & \quad 8715 \quad    & \quad 10120 \quad   & \quad 9205 \quad    \\
    \hline
  $\lambda^{\mathrm{g.s.}}_\mathrm{Exp}$ \quad        & \quad $6.45\cdot 10^{1}$ \quad  & \quad $4.09\cdot 10^{1}$ \quad  & \quad $1.12\cdot 10^{-1}$    \quad & \quad $7.45\cdot 10^{0}$ \quad \\
  $\Delta\lambda^{\mathrm{g.s.}}_\mathrm{Exp,Yield}$  & \quad $1.17\cdot 10^{1}$ \quad  & \quad $0.74\cdot 10^{1}$ \quad  & \quad $0.21\cdot 10^{-1}$    \quad & \quad $1.32\cdot 10^{0}$ \quad \\
  $\Delta\lambda^{\mathrm{g.s.}}_\mathrm{Exp,Approx}$ & \quad $1.39\cdot 10^{1}$ \quad  & \quad $0.99\cdot 10^{1}$ \quad  & \quad $0.34\cdot 10^{-1}$    \quad & \quad $0.46\cdot 10^{0}$ \quad \\
  $\lambda^{\mathrm{g.s.}}_\mathrm{TALYS}$   & \quad $8.19\cdot 10^{1}$ \quad  & \quad $7.74\cdot 10^{1}$ \quad  & \quad $2.07\cdot 10^{-1}$    \quad &  \quad $1.11\cdot 10^{1}$    \quad \\
  $\lambda^{\mathrm{g.s.}}_\mathrm{N.S.}$   & \quad $8.62\cdot 10^{1}$ \quad  & \quad $7.16\cdot 10^{1}$ \quad  & \quad $2.06\cdot 10^{-1}$    \quad &   \quad $9.21\cdot 10^{0}$    \quad \\
  \hline\hline
  $T=3.0\times 10^9$~K & & & & &\\
  Upper energy limit  & \quad 8945 \quad   & \quad 9010 \quad    & \quad 10400 \quad    &  \quad 9505 \quad    \\
  \hline
  $\lambda^{\mathrm{g.s.}}_\mathrm{Exp}$ \quad & \quad $2.92\cdot 10^{4}$ \quad  & \quad $1.78\cdot 10^{4}$ \quad  & \quad $1.49\cdot 10^{2}$    \quad &  \quad $5.04\cdot 10^{3}$    \quad \\
  $\Delta\lambda^{\mathrm{g.s.}}_\mathrm{Exp,Yield}$  & \quad $0.54\cdot 10^{4}$ \quad  & \quad $0.33\cdot 10^{4}$ \quad  & \quad $0.29\cdot 10^{2}$    \quad &  \quad $0.88\cdot 10^{3}$    \quad \\
  $\Delta\lambda^{\mathrm{g.s.}}_\mathrm{Exp,Approx}$  & \quad $0.99\cdot 10^{4}$ \quad  & \quad $0.67\cdot 10^{4}$ \quad  & \quad $0.52\cdot 10^{2}$    \quad & \quad $0.28\cdot 10^{3}$    \quad \\
  $\lambda^{\mathrm{g.s.}}_\mathrm{TALYS}$   & \quad $3.31\cdot 10^{4}$ \quad  & \quad $3.25\cdot 10^{4}$ \quad  & \quad $2.58\cdot 10^{2}$ \quad &  \quad $6.79\cdot 10^{3}$    \quad \\
  $\lambda^{\mathrm{g.s.}}_\mathrm{N.S.}$   & \quad $3.39\cdot 10^{4}$ \quad  & \quad $2.91\cdot 10^{4}$ \quad  & \quad $2.58\cdot 10^{2}$ \quad &  \quad $5.65\cdot 10^{3}$    \quad \\
  \hline\hline
  \end{tabular}

\caption{The ground-state reaction rates (in s$^{-1}$) determined
experimentally by the approximation approach of a thermal Planck
spectrum for three different temperatures are in good agreement
with two theoretical predictions using the TALYS and
NON-SMOKER$^{\mathrm{WEB}}$ code. The energy range between the
neutron emission threshold energy and the quoted upper energy
limit (in keV) is estimated to contribute about 99\% to the total
ground-state reaction rate.
$\Delta\lambda^{\mathrm{g.s.}}_\mathrm{Exp,Yield}$ and
$\Delta\lambda^{\mathrm{g.s.}}_\mathrm{Exp,Approx}$ account for
the uncertainty of the experimental reaction yield and for the
uncertainty stemming from the approximation of the Planck spectrum
within the Gamow-like window,
respectively.}\label{tab:ReactionRates_Results}
\end{table*}

\subsection{Determination of ground-state reaction rates}\label{sec:GroundStateReactionRates}
We have derived the $(\gamma,n)$ ground-state reaction rates  for
$^{148,150}$Nd, $^{154}$Gd, and $^{154}$Sm at temperatures between
2 and $3\times 10^9$~K using the approximation of a thermal Planck
spectrum as described in Sec.~\ref{sec:Reaction-Rates}. The
results are presented in Tab.~\ref{tab:ReactionRates_Results}.\\
The experimental uncertainties
$\Delta\lambda^{\mathrm{g.s.}}_\mathrm{Exp,Yield}$ and
$\Delta\lambda^{\mathrm{g.s.}}_\mathrm{Exp,Approx}$ stem from the
experimental uncertainty of the reaction yield determination and
from the uncertainty of the approximated thermal Planck spectrum
within the Gamow-like window, respectively. As will be discussed
in Sec.~\ref{sec:Error_discussion}, the experimental data points
close to the neutron emission threshold suffer from large
uncertainties. Hence, we discarded these data points for the
analysis to increase the reliability of the experimental
determination of the ground-state reaction rates, although the
approximation of the Planck spectrum, therefore, became less
accurate. Moreover, for the isotopes $^{148,150}$Nd and $^{154}$Gd
experimental data points were missing at energies further above
the neutron separation energy. This also reduced the accuracy of
the approximation of the Planck spectra at temperatures close to
$3\times 10^9$~K. Thus, except for the case of $^{154}$Sm the
approximation in this experiment was less accurate than stated in
Sec.~\ref{sec:Reaction-Rates}. For $^{160}$Gd too few data points
were available in the energy region of interest to derive ground-state reaction rates using the approximation approach.\\
For comparison the ground-state reaction rates were calculated
from Eq.~(\ref{eq:Reaction-Rate}) using the unnormalized
theoretical predictions for the photoneutron cross sections
discussed in the preceding section. As shown in
Tab.~\ref{tab:ReactionRates_Results} the calculations are in good
agreement with the experimental results within the experimental
uncertainties and the uncertainties of the approximation.
Consistent with the results found for the normalization factors of
the photoneutron cross section the statistical model calculations
tend to slightly overestimate the reaction rates for the studied
isotopes.

\subsection{Discussion of systematic and statistical uncertainties}\label{sec:Error_discussion}
An overview of the various experimental uncertainties is shown in
Tab.~\ref{tab:Errors}. We distinguish between systematic
 uncertainties $\Delta_\mathrm{sys}$ and statistical uncertainties
$\Delta_\mathrm{stat}$. The latter are not correlated between the
different experimental runs at various energies $E_\mathrm{max}$
and can, therefore, be reduced by a large number of individual
measurements. The contribution of the various uncertainties to the
overall uncertainty will be discussed in detail in the following.

\begin{table}[htbp]
  \centering
  \setlength{\extrarowheight}{2.pt}
\begin{tabular}{lcc}
\hline\hline
             & \quad $\Delta_{\mathrm{sys}}$ \quad & \quad $\Delta_{\mathrm{stat}}$\\
\hline
photon flux         \quad & \quad       16\%        \quad       &   \quad       5\%             \quad\\
$E_{\mathrm{max}}$  \quad & \quad       --           \quad       &   \quad       *              \quad\\
detection efficiency   \quad & \quad       4\%         \quad       &   \quad       2\%$^\dag$ / 15\%$^\ddag$      \quad\\
self-absorption     \quad & \quad       5\%      \quad       &   \quad       --              \quad\\
counting            \quad & \quad        --         \quad       &   \quad       2\%      \quad\\
$\gamma$-intensity  \quad & \quad       $3$\%      \quad       &   \quad       --              \quad\\
target mass         \quad & \quad       --          \quad       &   \quad       2\%      \quad\\
\hline total \quad     &  \quad   $18$\%      \quad & \quad
6\%$^\dag$ / 16\%$^\ddag$\\
\hline\hline
\end{tabular}
\caption{Systematic and statistical uncertainties of the
determination of the experimental $(\gamma,n)$ reaction yield. The
two uncertainties of the detection efficiency stated refer to the
HPGe setup ($^\dag$) and the LEPS setup ($^\ddag$), respectively.
The uncertainties stemming from an uncertainty in the maximum
photon energy $E_\mathrm{max}$ (*) are illustrated in
Fig.~\ref{fig:Yield_Uncertainty}. The total uncertainty is
calculated by a Gaussian error propagation.}\label{tab:Errors}
\end{table}

\begin{figure}[htbp]
  \includegraphics[width=1.\columnwidth]{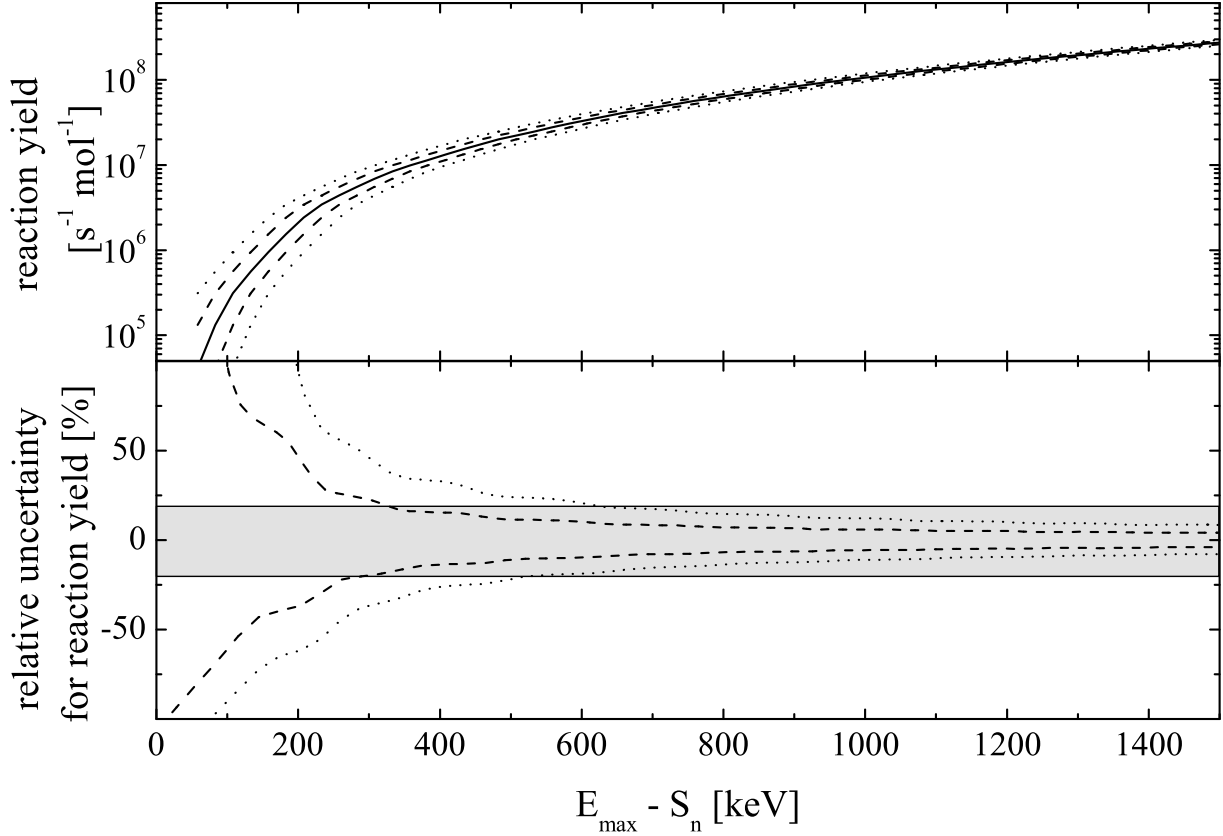}\\
  \caption{The expected $(\gamma,n)$ reaction yield of a $^{154}$Sm target (top panel) using a NON-SMOKER$^\mathrm{WEB}$ calculation
  for the photoneutron cross section (solid line). The dashed and dotted lines indicate the error bands assuming uncertainties of $E_\mathrm{max}$
  of 25~keV and 50~keV, respectively.  The bottom panel shows that close to the neutron threshold
this uncertainty clearly dominates the uncertainty stemming from
the photon flux determination (grey
band).}\label{fig:Yield_Uncertainty}
\end{figure}

The dominant uncertainty in our experiment comes from the
determination of the photon flux as discussed in
Sec.~\ref{sec:PhotonFlux}. The systematic uncertainty
$\Delta_{\mathrm{sys}}$ denotes the mean deviation between the
normalized {\sc Geant4} simulation and the experimental data
points stemming from the photoscattering reaction
$^{11}$B$(\gamma,\gamma')$. Therefore, $\Delta_{\mathrm{sys}}$
describes the uncertainty of the determined spectral distribution
of the photon flux. In addition $\Delta_{\mathrm{stat}}$ accounts
for the statistical uncertainties of the
$^{11}$B$(\gamma,\gamma')$
reaction yields used for the normalization of the simulation.\\
Close to the neutron emission threshold the reaction yield is
highly dependent on the maximum photon energy $E_{\max}$ of the
activation. Thus, small uncertainties of $E_{\max}$ close to the
neutron emission threshold give rise to large uncertainties in the
cross section determination as illustrated in
Fig.~\ref{fig:Yield_Uncertainty}. For our analysis we estimated
$E_{\max}$ to be known with an uncertainty of 25~keV. We assumed
that there was no systematic deviation of $E_{\max}$ inherent to
all experimental runs and, hence, no systematic uncertainty was
taken into account. Therefore, the uncertainty of $E_{\max}$ was
only treated statistically in the analysis. Due to averaging over
many individual measurements at various $E_{\max}$, the
uncertainty of $E_{\max}$ only represents a minor
contribution of the order of a few percent to the total uncertainty of the determined normalization factors.\\

To derive the reaction yield from the measured activities of the
produced unstable isotopes the detection efficiency has to be
known accurately. As already discussed in
Sec.~\ref{sec:Reaction_Yield_Determination} the detection
efficiency was determined with an uncertainty of 7\%. Since each
reaction yield was derived from several $\gamma$-transitions of
different energies the average systematic uncertainty was reduced
to about 4\%. Besides, it was found that the observed activation
count rate very sensitively depended on a proper placement of the
target in front of the detectors. Repeated measurements with
calibration sources showed that an additional uncertainty of 2\%
at the HPGe setup and, due to the very short distance between
target and detector, an uncertainty of 15\% at the LEPS setup had
to be taken into account. These measurements also proved that this
uncertainty could be treated statistically, and, hence, was
reduced by averaging over many data points.\\
Other uncertainties stemming from the self-absorption of emitted
photons within the target, the statistical uncertainty of the
counting, the $\gamma$-intensities of the decays and the target
mass were typical of the order of a few percent.\\
The total experimental uncertainty derived from the various
contributions discussed in this section is stated in
Tab.~\ref{tab:Errors}. As can be seen, the accuracy is mainly
limited by systematic uncertainties. Nevertheless, one also has to
account for large statistical uncertainties, when the reaction
yield is determined for a single experimental run. However, the
statistical uncertainties are significantly reduced when averaging
over many data points in the analysis as done for the deduction of
the normalization factors and ground-state reactions rates.
Therefore, the total experimental uncertainties for the results
presented in this work are typical of the order of 20\%.

\section{Summary and conclusion}\label{sec:Summary}
In the last years extensive studies of photoneutron reactions with
astrophysical implication for isotopes in the mass region $A\geq
186$ have been carried out, but no experimental data have been
provided for the rare earth isotopes so far. However, reliable
experimental data for a wide range of isotopes in various mass
regions are mandatory in order to constrain and to improve the
theoretical calculations required for astrophysical reaction
networks. The intention of this work was to test the theoretical
predictions of two different statistical model codes for the
photoneutron cross sections and ground-state reaction rates in the
rare earth region. Although a variety of photoneutron reactions in
this mass region was studied in the 70s, these experiments mainly
focused on the investigation of the Giant Dipole Resonance. Hence,
data points stemming from these experiments usually underlie large
statistical and systematical uncertainties close to the neutron
threshold and do not serve as a reliable test for theoretical
predictions in the astrophysically relevant energy region.
Therefore, we determined normalization factors for the
calculations of the photoneutron cross sections of the isotopes
$^{148,150}$Nd, $^{154}$Sm, and $^{154,160}$Gd in photoactivation
experiments close to the neutron separation energy. Moreover, we
derived $(\gamma,n)$ ground-state reaction rates in the
astrophysically relevant energy region for the \emph{p} process
using a superposition of bremsstrahlung spectra of various energies.\\
The need of experimental data of photoneutron reactions for
astrophysical reaction networks is twofold. First, for
\emph{p}-process studies reliable predictions of reaction rates
for a wide range of isotopes need to be provided. Since most of
the isotopes involved in the \emph{p}-process reaction network are
unstable, the nuclear physics input for these isotopes is rare
and, hence, theoretical calculations usually cannot be adjusted
locally. Therefore, the reliability of the theoretical models
should be tested using only global input parameters. In this work
we found that in the rare earth region the predicted cross
sections and the reaction rates for the selected isotopes agree
with the experimental data within a factor of two. On average, a
mean uncertainty of less than 30\% has been determined. This is
consistent with the results found in previous experiments in the
mass region $A\geq 186$. In addition, no systematic dependence of
the deviation with atomic mass has been observed. On the one hand,
this proves the reliability of the statistical model codes for a
wide range of isotopes along the valley of stability. But it also
indicates an appropriate predictive power for the reaction rates
of unstable isotopes, since only global input parameters were used
in the calculations . However, for almost all studied isotopes the
cross sections were found to be overestimated by theory, which
calls for further investigations. Moreover, it is planned to study
the isotopic chain of the cerium isotopes in the near future to
test the validity of the statistical model in this mass region for
isotopes close to and at the neutron shell closure. Although the
predictions agree fairly well with the experimental data of this
work, it needs to be emphasized that experimental studies cannot
account for the stellar enhancement of cross sections and reaction
rates due to the thermal population of low-lying levels in the
parent nuclei under stellar conditions. This so-called
\emph{stellar enhancement
factor} can only be derived from theory.\\
The second astrophysical application of experimental studies of
photoneutron reactions arises in the investigation of branching
points within the \emph{s}~process. In this context the
photoneutron reactions of $^{148}$Nd and $^{154}$Gd are of
particular interest among the studied isotopes of this work. Since
the neighbouring branching points $^{147}$Nd and $^{153}$Gd cannot
be studied in neutron capture experiments due to their short
half-life, the aim is to derive the neutron capture cross sections
of these branching points from the photoneutron reaction of the
isotopes $^{148}$Nd and $^{154}$Gd, respectively. It was shown in
this work, that the two adopted theoretical predictions
overestimate the photoneutron cross sections by up to a factor of
two. This might indicate that also the predicted neutron capture
cross section of the inverse reactions need to be adjusted
accordingly when being calculated within the same statistical
model code. For astrophysical studies, a direct correlation
between the stellar neutron capture and photodisintegration rate
is described by the so-called \emph{principle of detailed balance}
\cite{Fowl67,Raus00a}. This correlation only holds under the
assumption that the low-lying levels in both the parent and the
residual nucleus are thermally populated to a significant amount.
This condition is fulfilled under sufficiently hot temperatures in
a stellar environment. However, experiments in the laboratory only
address transitions starting from the ground state of a certain
nucleus and, therefore, the principle of detailed balance is not
applicable any more for a direct comparison of the laboratory
neutron capture and photodisintegration rate. Hence, without any
further assumptions, conclusions for the neutron capture rate can
hardly be drawn from the reaction rates derived in
photodisintegration
experiments.\\
Instead of directly deriving the neutron capture reaction rate
from experiment a promising approach is to improve the nuclear
physics input of the statistical model codes in order to increase
the reliability of the theoretical predictions. To obtain the most
accurate calculations for a single nucleus the input parameters
should be adjusted locally. Therefore, the aim of future studies
should be to provide improved nuclear physics input for $^{148}$Nd
and $^{154}$Gd, while the experimental results of this work serve
as a reliable test for any new complete set of input parameters.
It can be assumed that a set of input parameters yielding improved
predictions for the photoneutron reaction, will then also enhance
the calculations of the neutron capture cross sections of the
branching points $^{147}$Nd and $^{153}$Gd. At this point,
however, we would like to make clear that the experimental data
provided by this work set constraints on the absolute value of the
photoneutron cross section, but do not allow for separately
adjusting the various nuclear physics parameters such as the
neutron optical potential, nuclear level densities and the
$\gamma$-ray strength function. Therefore, further experimental
investigations with direct access to these parameters are
mandatory in the future.

\section{Acknowledgments}
We thank the S-DALINAC group around R.~Eichhorn for the reliable
beam during our experiment and T.~Rauscher for his support with
the theoretical calculations using the NON-SMOKER$^{\mathrm{WEB}}$
code. Moreover, we thank the Karlsruhe group around F.~K\"appeler
for helping us with the target preparation. This work was
supported by the Deutsche Forschungsgemeinschaft under contract
SFB 634.

\newpage 

\end{document}